\documentclass[fleqn,usenatbib]{mnras}

\usepackage{newtxtext,newtxmath}
\usepackage[T1]{fontenc}
\usepackage{ae,aecompl}

\usepackage{graphicx}	
\usepackage{amsmath}	
\usepackage{multirow}
\usepackage{array}
\usepackage[export]{adjustbox}
\usepackage{rotating}
\usepackage[utf8]{inputenc}

\usepackage{tikz}
\usepackage{collcell}
\newcommand*{\MinNumber}{-1.0}%
\newcommand*{\MidNumber}{0.0} %
\newcommand*{\MaxNumber}{1.0}%
\definecolor{amber}{rgb}{1.0, 0.49, 0.0}
\definecolor{aqua}{rgb}{0.0, 1.0, 1.0}
\def \ifempty#1{\def\temp{#1} \ifx\temp\empty }
\newcommand{\ApplyGradient}[1]{%
    \ifempty{#1}
        #1
    \else
        \ifdim #1 pt > \MidNumber pt
            \pgfmathsetmacro{\PercentColor}%
                {max(min(100.0*(#1-\MidNumber)/(\MaxNumber-\MidNumber),100.0),0.00)}%
            \hspace{-0.95mm}\colorbox{amber!\PercentColor!white}{#1}
        \else
            \pgfmathsetmacro{\PercentColor}%
                {max(min(100.0*(\MidNumber-#1)/(\MidNumber-\MinNumber),100.0),0.00)}%
            \hspace{-0.95mm}\colorbox{white!\PercentColor!aqua}{#1}
        \fi
    \fi
}
\newcolumntype{Z}{>{\collectcell\ApplyGradient}c<{\endcollectcell}}
\newcolumntype{P}[1]{>{\centering\arraybackslash}p{#1}}
\newcolumntype{C}[1]{>{\centering\arraybackslash}m{#1}}
\newcolumntype{R}[1]{>{\raggedleft\arraybackslash}p{#1}}

\title[Glob. Weather for Astronomical Obs.]{Global Weather for the Astronomical Observatories}

\author[Z. Kurt et al.]{%
Z. Kurt$^{1,2}$\thanks{E-mail:zuhal.kurt1990@gmail.com},
S. K. Yerli$^{3}$,
N. Aksaker$^{4,2}$,
A. Aktay$^{5,1}$, 
M. Bayazit$^{1}$,\newauthor
M.A. Erdoğan$^{6}$,
\\
$^1$Remote Sensing and Geographic Information System, University of Cukurova, 01330 Adana, Turkey\\
$^2$Space Science and Solar Energy Research and Application Center (UZAYMER), University of Çukurova, 01330 Adana, Turkey\\
$^3$Department of Physics, Orta Doğu Teknik Üniversitesi, 06800 Ankara, Turkey\\
$^4$Adana Organised Industrial Zones Vocational School of Technical Science, University of Çukurova, 01410 Adana, Turkey\\
$^5$Turkey State Meteorological Service, Regional Forecast Center, 01360 Adana, Turkey \\
$^6$Landscape Architecture Department, Faculty of Architecture, Hatay Mustafa Kemal University, 31060 Hatay, Turkey.\\
}

\date{Accepted XXX. Received YYY; in original form ZZZ}
\pubyear{2021}

\begin{document}
\label{firstpage}
\pagerange{\pageref{firstpage}--\pageref{lastpage}}
\maketitle

\begin{abstract}
Astronomical sites occupying observing instruments have to be selected according to many factors.
Among these factors, geographic location of the site and quality of atmosphere above the site play an important role in the decision process.
The following factors were chosen to create layers at their geographic locations (observatories: 1905 northern, 235 southern) from the \href{https://www.astrogis.org}{astroGIS database}: CC (cloud coverage), PWV (precipitable water vapor), AOD (atmospheric optical depth), VWV (vertical wind velocity) and HWV (horizontal wind velocity).
In order to estimate astronomical importance of geographic location of the sites and quality of airmass above the sites, DEM (digital elevation model) and LAT (latitude of observatory location) layers were also included.
Two periodic variations have been produced from these factors: monthly and annual averages.
In addition to the variations or trends a complete statistical analysis was carried out for all factors to investigate the potential correlations between the factors:
There is a clear difference between northern and southern hemispheres.
Exchange of meteorological seasons between hemispheres are also compliant within factors.
The geographical locations of most of the observatories found to be ``not suitable'': On the average, DEM is low (550 m), CC is high (70\%) and PWV is high (14 mm).
There seems to be no apparent long-term variations and/or patterns in all factors.
We once again confirm the common expectation of astronomy: \textit{as DEM increases astronomical conditions get better} (CC, PWV and AOD gets lower values).
All the results will be made available online through astroGIS database.
\end{abstract}


\section{Introduction}
\label{sec:Introduction}

\begin{figure*}
\begin{center}
\includegraphics[width=\columnwidth]{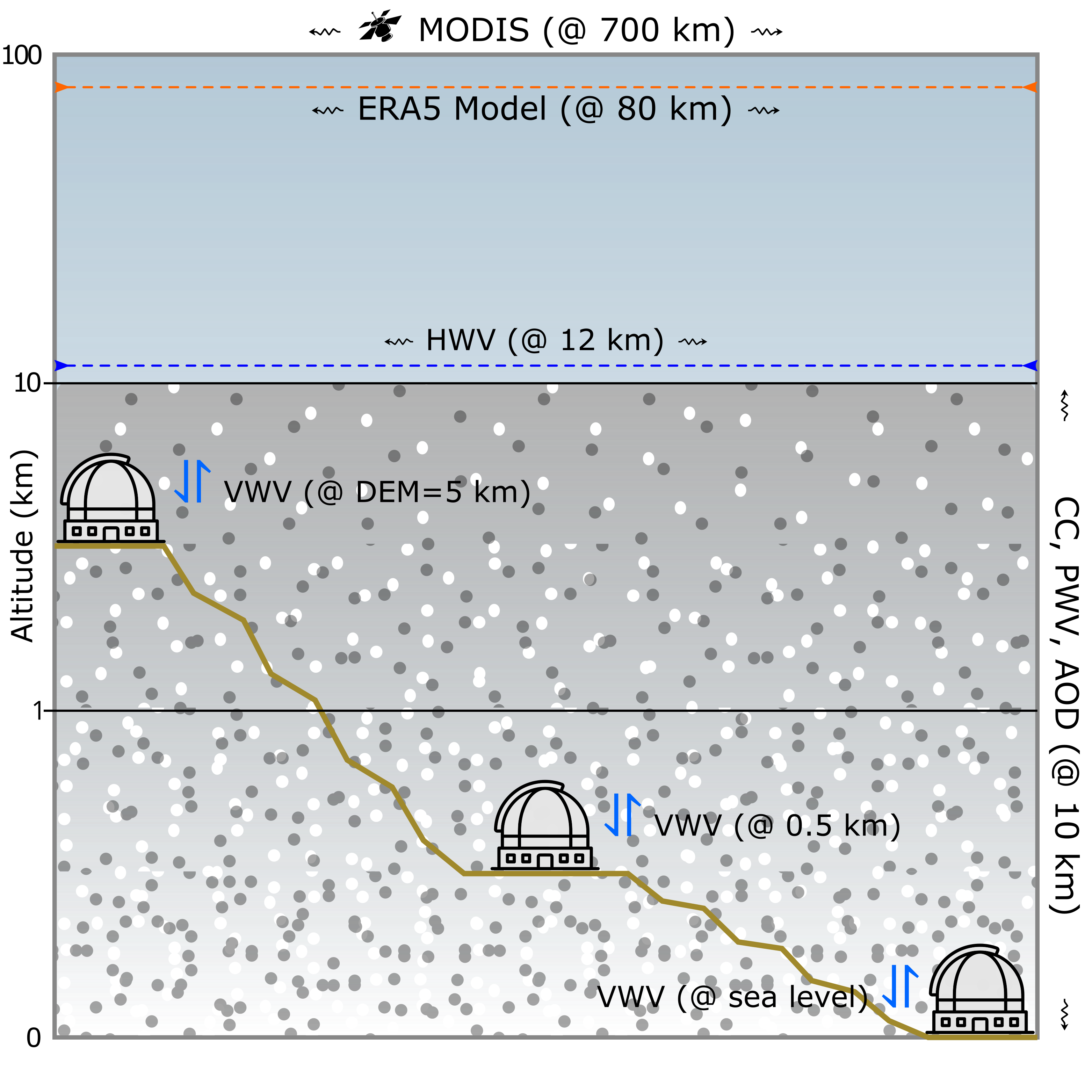}
\caption{%
Schematic view of Earth's lower atmosphere.
It shows the atmosphere up to the Karman line (100 km) in logarithmic scale.
All factors (DEM, CC, PWV, AOD, HWV and VWV) are marked from the astronomical point of view (see \S\ref{sec:factors} for their definitions); altitude, however, represents the DEM. 
Densities of CC (gray-to-white shading), PWV (gray dots) and AOD (white dots) increases as altitude reaches to the sea level.
Effect of VWV is divided into three altitudes: Highest, average and sea level.
Observatory icon was taken from \href{https://www.freepik.com}{Freepik}%
}
\label{F:atm}
\end{center}
\end{figure*}

The quality of the observations carried at ground based observatories depend on how astronomers identify the Earth's atmosphere and it's effects on the light passing through it \citep{2006Msngr.125...44S}.
The atmosphere is constantly observed and monitored, therefore, we have collected quite a lot of data which lead us to understand its behaviour and predict the dynamics of the atmosphere.
Since the ground based observatories are located on different geographic locations at different elevations, the atmospheric column over the observatories (local atmosphere) will vary and therefore, creating associated variation in the quality of the observations.
However, because of the global variations caused by both the solar radiation and human activity on the ground, the local atmosphere will also show long-term variations.
Therefore, astronomical sites have to be selected taking into account all abovementioned factors.
The key factors in the atmosphere that obstruct the incoming radiation from space are shown in Figure \ref{F:atm}. 
Incoming radiation is mainly obstructed by water droplets, ice crystals, dust and sea salt which can be identified as cloud formation.
Depending on the density of the clouds, they partially or totally block the observing window of the ground based observatories.
Therefore, clouds are counted as the main factor that affect the ground based observations.
When water based cloud content is combined with the humidity (a.k.a density of water vapor) in the lower atmosphere, Precipitable Water Vapor (PWV) is said to be formed.
It is defined as column density of condensed water normal to the observing location in units of mm.
While clouds affects whole optical window from UV to IR, PWV mainly disturbs the radiation including and above IR wavelengths \citep{2013Msngr.152...17S, 2015MNRAS.452.1992P}.
In the lowest part of the atmosphere (around 10 km above the ground), however, smaller sized particles ($< 10 \mu$m) will affect the radiation and decrease the atmospheric transparency due the Mie Scattering \citep{1908AnP...330..377M} creating an atmospheric extinction \citep{2004SPIE.5489..138S}.
This is defined as the Aerosol Optical Depth (AOD).
These atmospheric abundances (clouds, PWV and AOD) are counted as the main catalyst in meteorological events happening in the troposphere \citep{2008MNRAS.391..507V}.
Weather in the troposphere is created due to dynamical fluctuations of atmospheric abundance which is defined as the wind; the energy is exchanged between local parts of the atmosphere.
Therefore, the incoming radiation reaching to the ground will be affected by wind speed at 200 hPa, corresponding to atmospheric volume around 12 km above ground \citep{2019MNRAS.482.4941H}.
In addition to this, horizontal wind velocity at the ground level which is associated with the orography of the surface (i.e Digital Elevation Model - DEM), will also affect and disturb the incoming radiation.
The wind is also related with the seeing of the astronomical images in observatories \citep{2007MNRAS.381.1179T}. 
The effects of these meteorological parameters on astronomical observations were shown by \cite{2007PASP..119.1186S, 2008BASI...36..111S, 2011PASP..123.1334V}. 
In summary, the light will be disturbed by clouds, PWV, AOD and wind profiles, and creating a deformed radiation at the ground level which is defined as the seeing of the location at the moment of the observation.

\begin{figure*}
\begin{center}
\includegraphics[height=\columnwidth]{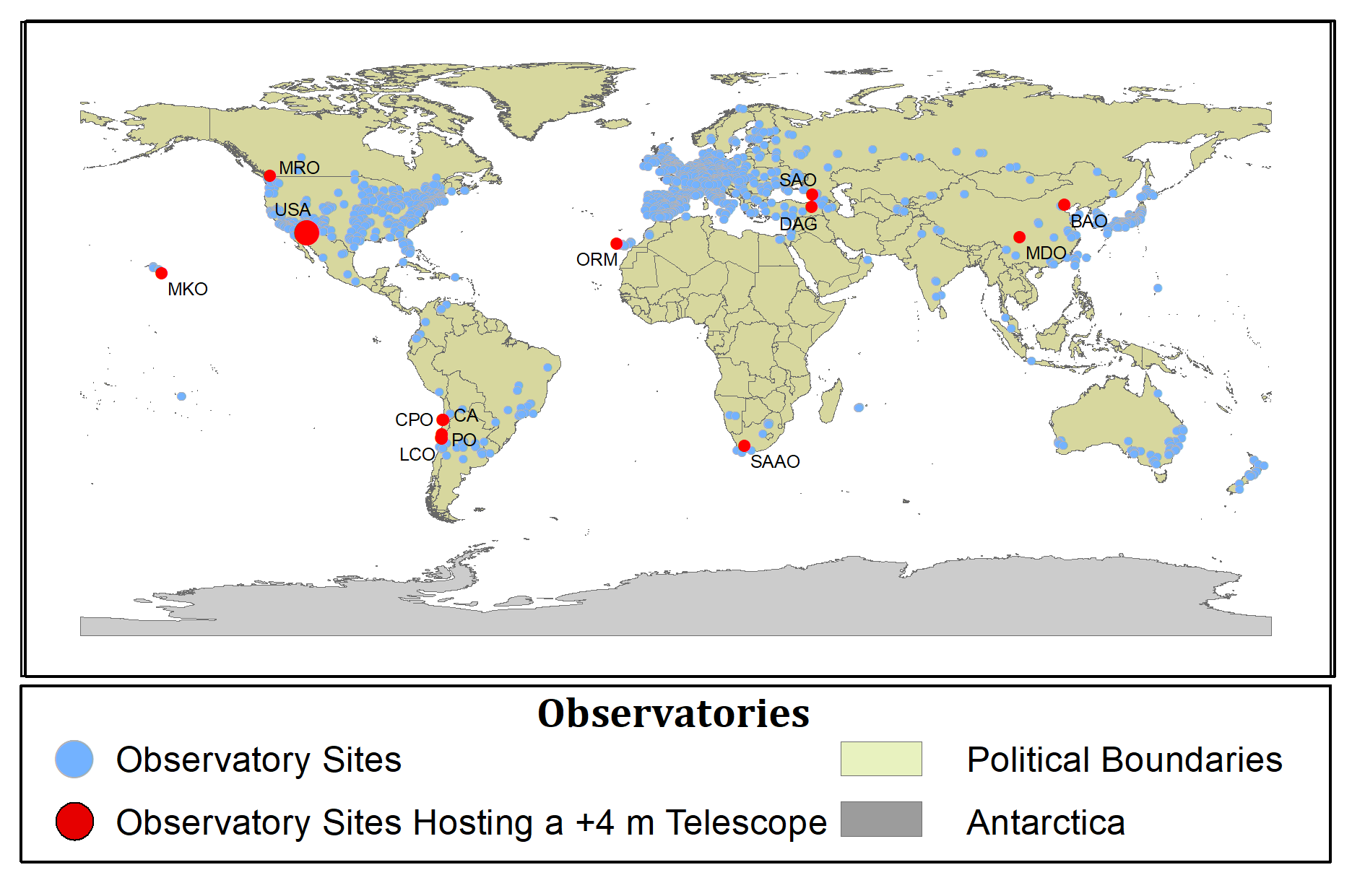}
\caption{%
Global distributions of observatories.
Note that observatories labelled as FLWO, KPNO, MGIO, PalO and LO are marked with a single wide red circle.
Observatories are shown with the blue dots. The red dots show observatories with +4m telescope. 
Since astroGIS doesn't include Antarctica, observatories and/or facilities in this continent are also excluded.
}
\label{F:obs}
\end{center}
\end{figure*}

Therefore, a global weather for astronomical observatories could only be studied by monitoring abovementioned factors from space (see Fig. \ref{F:obs} for a global distribution of all observatories).
This is due to the fact that space observations have a global coverage and high resolution of data.
In an earlier global study, \cite{2020MNRAS.493.1204A} collected a series of satellite data corresponding to different meteorological and astronomical parameters named as ``layers''.
They have released the dataset as \href{https://www.astrogis.org}{astroGIS}.
Using the astroGIS datasets they have created an index describing up-to-date quality of each observing site with a resolution of 1 km.
There are four SIAS (Suitability Index for Astronomical Sites) series describing different weights of data layers where each index ranges from 0 to 1.
Since global wind data is a model created by analysing the whole atmospheric data \citep[i.e. ERA5:][]{2020QJRMS.146.1999H}, it has to be integrated into astroGIS database which is completed in this study.

The long-term variations in the local weather above the observatories have to be forecast to create a profile of site's observing quality.
This can be achieved by adding time, including the seasonal variations, as a parameter to the astroGIS database.
Therefore, finding correlations between factors affecting the seeing (clouds, PWV, AOD and wind profiles) is the key in understanding the behavior of the local weather of the observatories.
The aim of this study is to produce these correlations and find out short- and long-term variations and trends for all observatories in a time span of 2000--2019.

\section{Databases}
\subsection{astroGIS}
The astroGIS database contains up-to-date and high resolution ($\sim$ 1 km) satellite data for six layers \citep{2020MNRAS.493.1204A}; click the database \href{https://www.astrogis.org/}{link} for further details.
The main properties of these layers are given in Table \ref{T:astrogis}.
The total database size is approximately 100 TB.
Datasets named as Cloud Cover (CC), Precipitable Water Vapor (PWV), Aerosol Optical Depth (AOD) were obtained from MODIS (Moderate Resolution Imaging Spectroradiometer) instrument in polar orbiting twin satellites, Aqua and Terra.
The astroGIS database also contains DEM which was obtained from GTOPO30.
The global geographic coverage of the database is from 75 North to 65 South.

The details of datasets (CC, PWV, AOD) in the astroGIS database are explained in \cite{2020MNRAS.493.1204A}, and their original data are collected from the MODIS instrument.
Note that, analysis of the CC in the astroGIS database is the most comprehensive one with the highest spatial resolution ever used in the cloud research.

\subsection{Wind Model -- ERA5}
European Centre for Medium-Range Weather Forecasts (ECMWF) provides full-time operational service and produces digital weather forecasts and other data.
They have one of the largest meteorological data archives in the world.
They periodically use predictive models and data implementation systems to re-analyze the archived observations to produce a final version of ERA5 wind model \citep{2020QJRMS.146.1999H}.
Estimates are produced for all locations around the world for a time span starting from year 1979.
The resolution of the ERA5 atmosphere and terrain re-analysis is 31 km.
The atmospheric component consists of 137 vertical levels up to 1 Pa (approximately 80 km) from the surface.
Due to the high temporal resolution the dataset contains many climatic statistics \citep{2020IJCli..40..979M}.

Since the wind speed at around 12 km which corresponds to 200-hPa pressure level, contributes to the astronomical seeing, a horizontal component of wind, namely $u_{200}$, will be calculated and used as the Horizontal Wind Velocity (HWV) from the ERA5 model.
Similarly, to be able to reflect the stability or instability of the airmass, Vertical Wind Velocity (VWV) is also taken from the ERA5 model.
VWV and HWV datasets are downloaded from \href{https://cds.climate.copernicus.eu}{ERA5 database}.
For each night ERA5 gives 8 points (3 before midnight, 1 at midnight and 4 after midnight) taken hourly.
A monthly averaged data is then created and used in this study.
\section{Analysis}
\label{sec:analy}

\subsection{The Factors}
\label{sec:factors}
The astroGIS database is retrieved and stored in the instrument's original format.
However, to have a good sense of global weather profile they have to have a temporal binning for related atmospheric variations.
In the first run on the data, all factors have been binned into daily averages.
Note that, these daily averages on the factors affecting the atmosphere will produce statistically irrelevant profiles, therefore, they are further binned into monthly (calendar month) and annual averages.
Moreover, since the time span of the whole dataset is long enough, weekly averages (52 points per year) would probably give more detailed variation profile on each factor.
However, to be able to compare results of this work with the literature, monthly (12 points per year) and annual binning strategy have been implemented.
Monthly and annual binning will give the trend of the seasonal and long-term profiles of the observatories, respectively.
The datasets are produced from astroGIS database using {\scshape Python} codes written in-house.

Implementation of each factor into the global analysis of the weather are given in the following paragraphs.
For CC, PWV, AOD and DEM pixel values of the observatories were read from the daily GEOTIFF images obtained from the astroGIS database.

\textbf{Cloud Coverage - CC}:
The data for this factor consist binary values \citep{2020MNRAS.493.1204A}.
The four binary values (0, 1, 2, 3) give four cloud classifications, namely ``Confident Clear'', ``Probably Clear'', ``Uncertain Clear'', ``Cloudy''.
The factor CC, therefore, can be constructed by taking into account only ``Cloudy'' classification while taking the averages over time (monthly and annual) creating the CC percentile, which will answer this question: \textit{how cloudy the site is}.

\textbf{Precipitable Water Vapor - PWV}, \textbf{Aerosol Optical Depth - AOD} and \textbf{Digital Elevation Model - DEM}:
Analysis of these three factors are similar to CC.
However, original data contains directly the value with their original units at the geographic location.

\textbf{200-hPa Wind Velocity - HWV}: 
Horizontal Wind Velocity (HWV) contains two velocity components: V (horizontal, moving to north) and U (horizontal, moving to east) \citep{2016AGUFMNG33D..01H}.
Therefore HWV vector has been calculated from these components.
Both monthly and annual averages are then calculated.
VWV data reduction procedure has been applied to HWV.

\textbf{Vertical Wind Velocity - VWV}:
NetCDF format of ERA5 database was retrieved for astroGIS database and they were reduced with several different official Python libraries.
VWV represents vertical (up: positive, down: negative) motion of air mass at the ground level.
The ECMWF data uses a pressure based vertical coordinate system.
Therefore it is combined with DEM data for each observatory, creating altitude correlated vertical wind data (VWV).

\subsection{Averages of the factors}

\begin{figure*}
    \centering
    \begin{tabular}{@{}c@{}c@{}}
    \includegraphics[height=4.25cm]{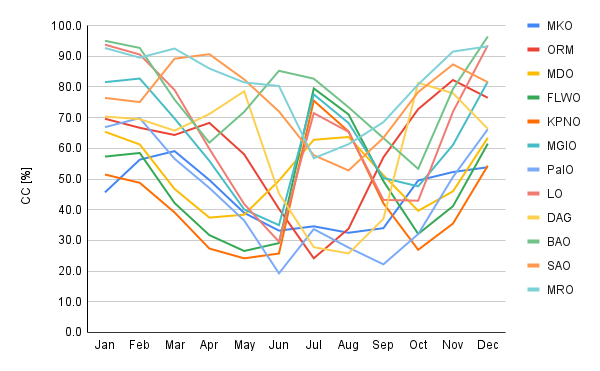}  &
    \includegraphics[height=4.25cm]{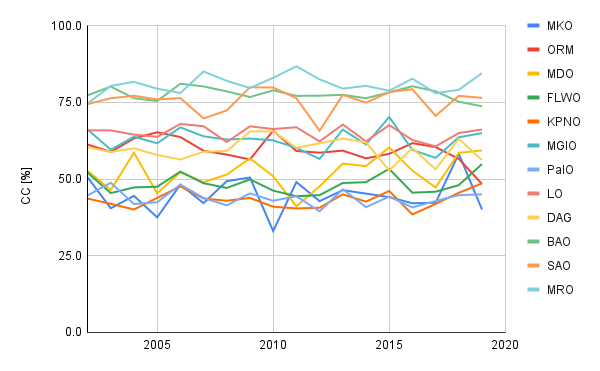}  \\
    \includegraphics[height=4.25cm]{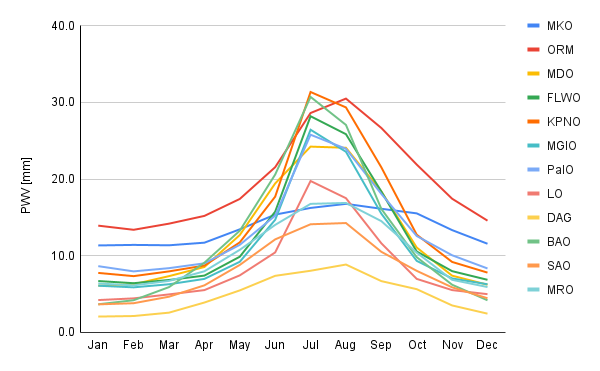} &
    \includegraphics[height=4.25cm]{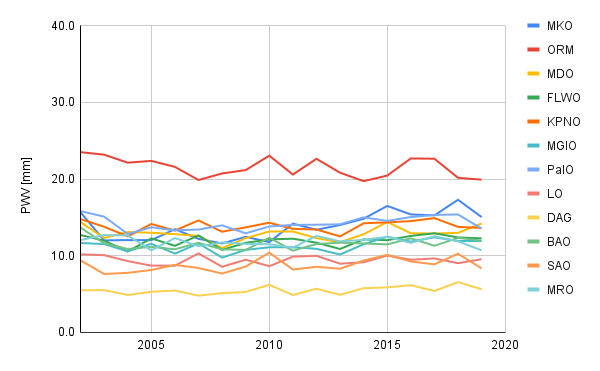} \\
    \includegraphics[height=4.25cm]{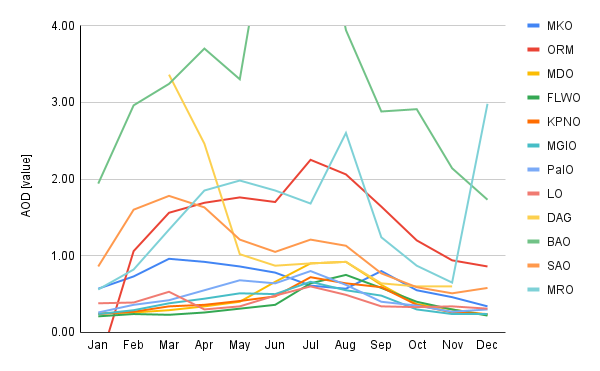} &
    \includegraphics[height=4.25cm]{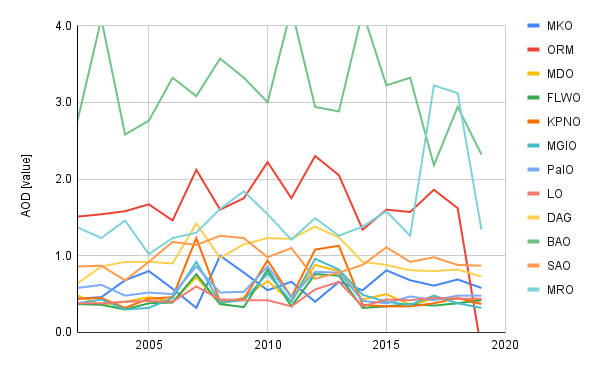} \\
    \includegraphics[height=4.25cm]{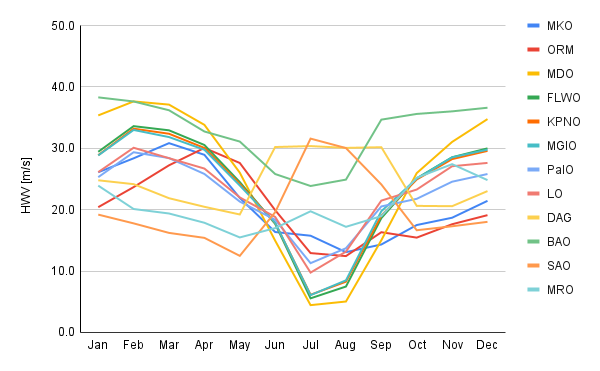} &
    \includegraphics[height=4.25cm]{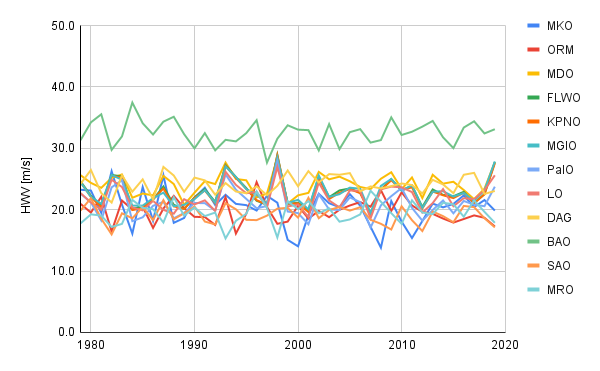} \\
    \includegraphics[height=4.25cm]{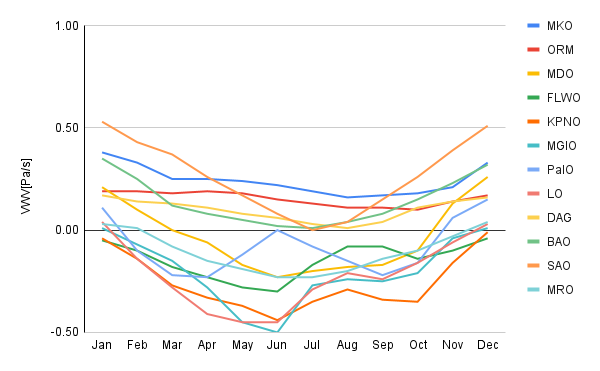} &
    \includegraphics[height=4.25cm]{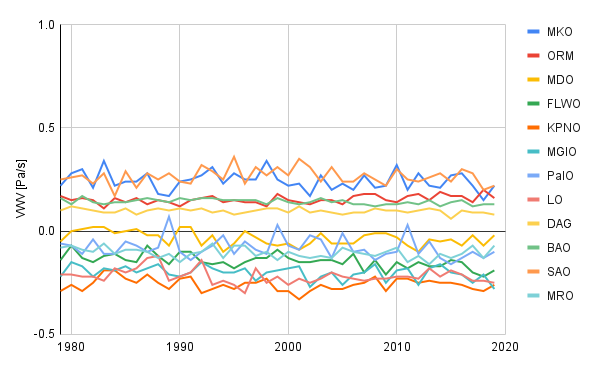} \\
    \end{tabular}
\caption{%
Monthly (left panel) and annual (right panel) averages for northern observatories for layers CC, PWV, AOD, HWV and VWV, from top to bottom, respectively.%
}
\label{F:north}
\end{figure*}

\begin{figure*}
    \centering
    \begin{tabular}{@{}c@{}c@{}}
    \includegraphics[height=4.25cm]{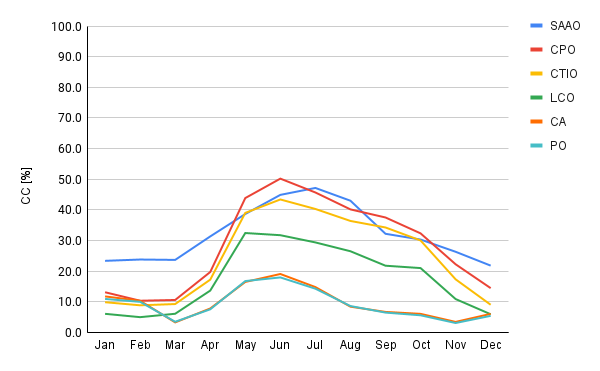}  &
    \includegraphics[height=4.25cm]{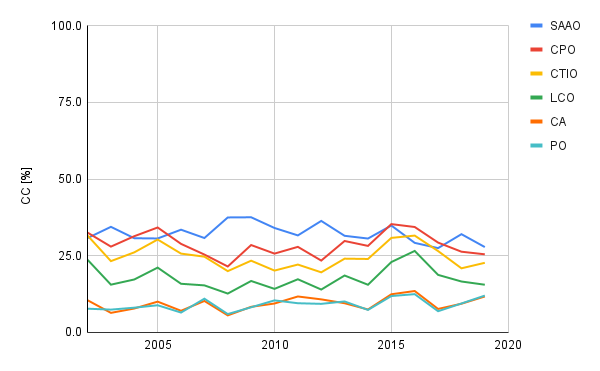}  \\
    \includegraphics[height=4.25cm]{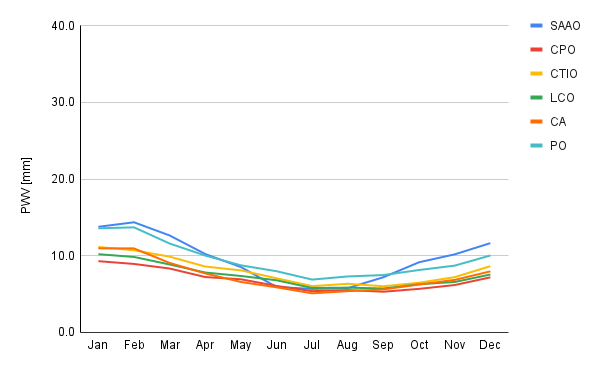} &
    \includegraphics[height=4.25cm]{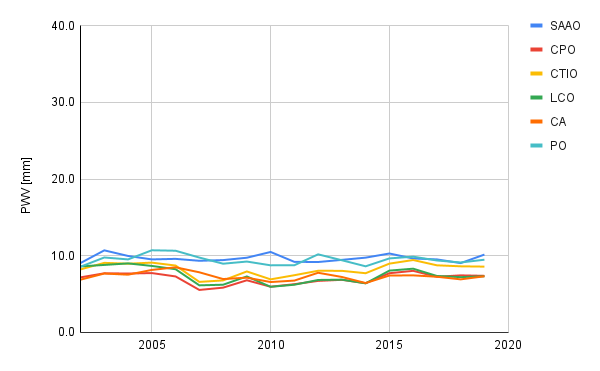} \\
    \includegraphics[height=4.25cm]{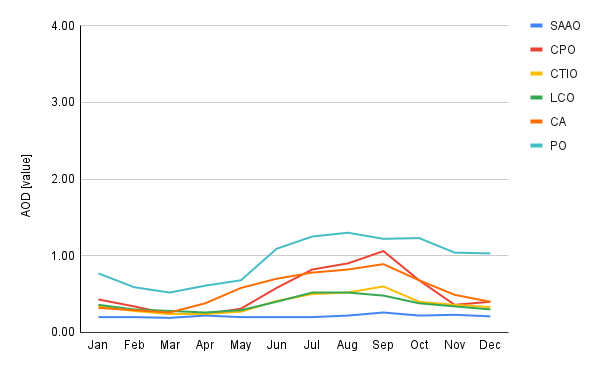} &
    \includegraphics[height=4.25cm]{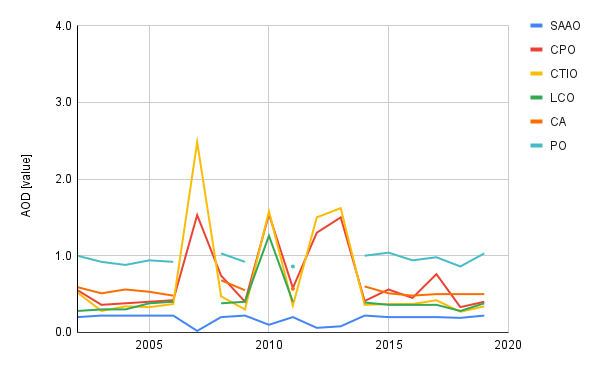} \\
    \includegraphics[height=4.25cm]{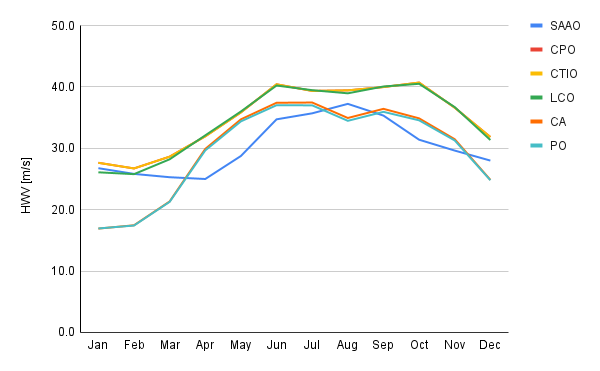} &
    \includegraphics[height=4.25cm]{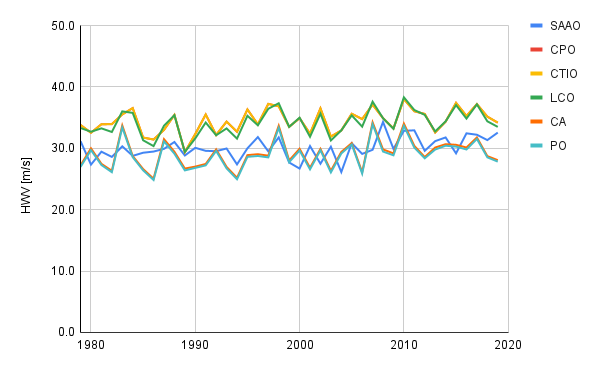} \\
    \includegraphics[height=4.25cm]{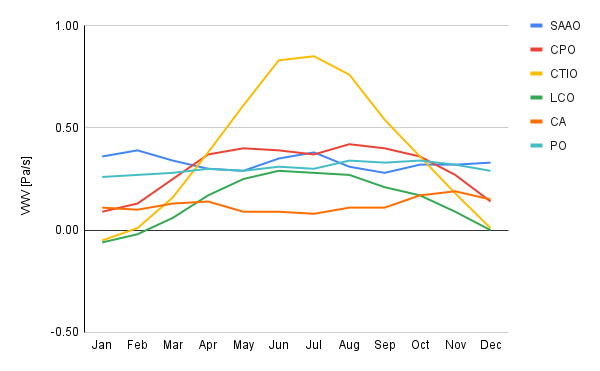} &
    \includegraphics[height=4.25cm]{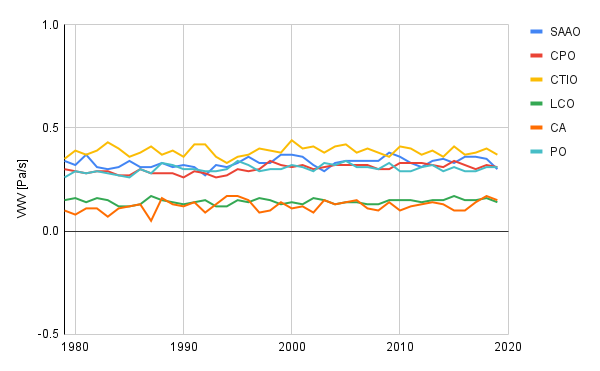} \\
    \end{tabular}
\caption{%
Monthly (left panel) and annual (right panel) averages for southern observatories for layers CC, PWV, AOD, HWV and VWV, from top to bottom, respectively.%
}
\label{F:south}
\end{figure*}

In the analysis of astroGIS database, average of annual variations (19 points for the time span of 2000--2019) will only give a single value for the site.
This will represent the level of site's corresponding factor independent from the time, and it does \textbf{not} represent the quality of the site.
In order to produce a global sense of local weather for a list of observatories, histograms of the factors over these annual average have to be created.
These histograms show data limits, number distributions and trends of these factors:
The histograms are given in Fig. \ref{F:hist}, and their statistics are tabulated in Table \ref{T:layer_stat}. 
Due to the global weather cycles which are divided mainly by hemispheres, the observatories falling on two different hemispheres are given in separate figures: Fig. \ref{F:north} (northern) and Fig. \ref{F:south} (southern).
In these figures, monthly and annual averages are given in left panel and right panel, respectively, for CC, PWV, AOD, VWV and HWV factors.
Note that, these two figures contain only 14 observatories having +4 m telescope size where their properties are given in Table \ref{T:obs}.
The averages for all observatories will be made available as a supplement to the article. 

\subsection{Trends of the factors}
\label{sec:Trends}

The stability of the observing quality is one of the main concern for each mid- or large-scale observatory.
These facilities try to estimate and/or forecast the local weather to maximize astronomical observing window (both spectrum and time wise).
In order to achieve this globally, linear regression fits were applied to annual averages of the factors described above on the long-term dataset in the astroGIS database.
As can be easily deduced from the figure, majority of the trends shows a variation almost stable around the average for each plotted observatory.

\subsection{Correlations of the factors}
\label{sec:Correlations}

Some of the well-known expressions are used to express the correlations among astrometeorological parameters.
The statement ``\textit{air, therefore sky, is clear in higher elevations}'' is taken for granted by the astronomers because (a) convective energy exchange (i.e. weather) mostly occurs in lower atmosphere; (b) therefore there will be less number of water molecules in higher elevations; (c) and since wind gets lower values as elevation increases, cloud formation decreases in higher elevations.
However, locations of observatories weren't chosen with respect to this simple reasoning and therefore our observatory list creates a good statistical distribution:
The list contains both low and high elevations, and observatory locations (longitude and latitude) are distributed in both north and south hemispheres, creating the required number statistics to be tested by the well-known statistical methods.

Annual averages are used in all three methods: Spearman ($\rho$), Pearson ($r$), Kendall($\tau$).
DEM and observatory latitude were added to the parameter space and the complete statistical analysis is given in Table \ref{T:coef}. 
Since our data is not categorized, Pearson and Kendall methods become not suitable to drive conclusions even though all three methods give similar trends in the analysis.
However for completeness they are all kept in the Table but only Spearman's values are used in the analysis.
Note that, interpretation of correlation value calculated between two variables is described as follows: -1: positive correlation, +1: negative correlation, close to 0: no correlation.
Therefore, to decide whether there is a correlation among two variables or not, the absolute value of the coefficient should be $>0.5$.
All the probabilities of the coefficients are found to less than 5\% therefore not tabulated in the table.

The results of the analysis (highest correlation first) are as follows.
\textbf{Correlations:}
DEM decreases with increasing VWV (-0.89);
LAT decreases with increasing HWV (-0.70);
DEM decreases with increasing AOD (-0.66);
VWV increases with increasing AOD (0.64);
AOD increases with increasing CC (0.57);
LAT increases with increasing CC (0.54).
\textbf{No Correlations:}
PWV and CC (-0.06);
DEM and HWV (0.08).
These six correlations, one of two no-correlations and a correlation candidate (PWV and DEM) are then plotted in Fig. \ref{F:corr_new} in the above order.
\section{Statistical Results}

\begin{figure*}
    \centering
    \includegraphics[width=0.49\textwidth]{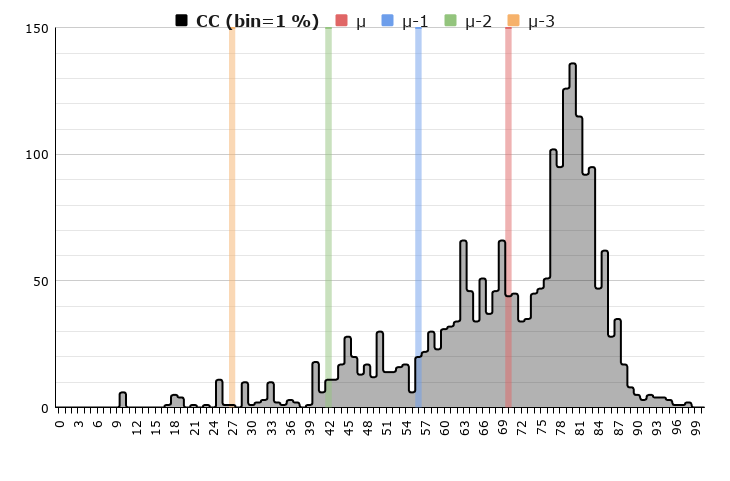}   
    \includegraphics[width=0.49\textwidth]{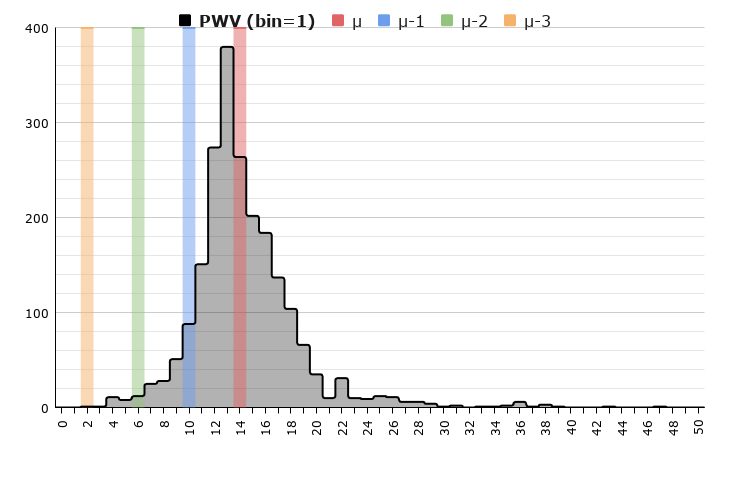}    
    \includegraphics[width=0.49\textwidth]{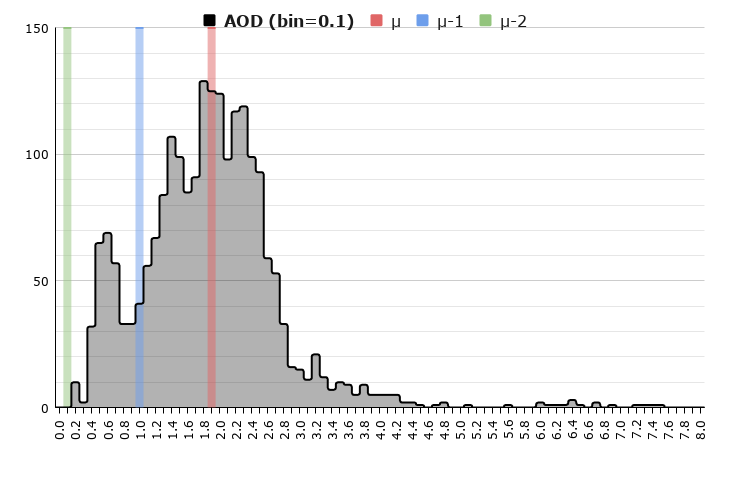}    
    \includegraphics[width=0.49\textwidth]{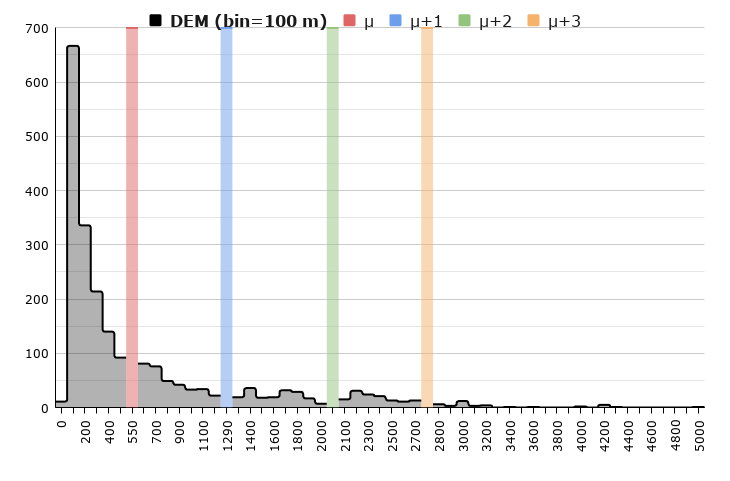}   
    \includegraphics[width=0.49\textwidth]{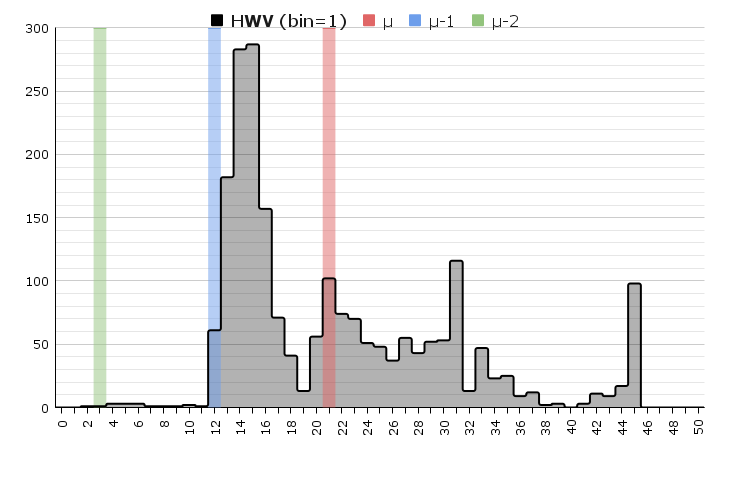}   
    \includegraphics[width=0.49\textwidth]{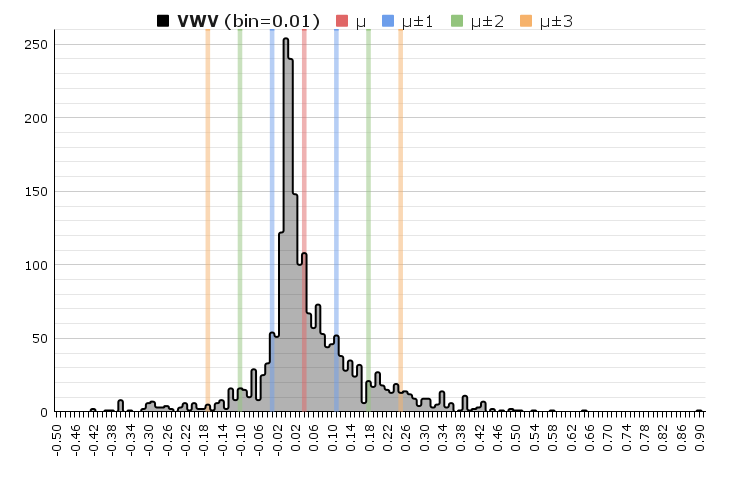}   
\caption{%
    Histograms of annual averages for observatories for the following factors:
    CC (upper left), PWV (upper right), AOD (middle left), DEM (middle right), VWV (bottom left), HWV (bottom right).
    Bin size is given upper right corner of the plot. 
    Except for DEM and VWV, the histograms include the mean $\mu$ (red bar), $\mu-1\sigma$ (blue bar), $\mu-2\sigma$ (green bar), $\mu-3\sigma$ (orange bar).%
}
\label{F:hist}
\end{figure*}

In this section, main discussion of statistical analysis will be given for each panel of Fig. \ref{F:hist} and Tables \ref{T:layer_stat} and \ref{T:factor_ave}.

List of the observatories used in this work as a database is far from a professional usage: Observatories producing scientific data are mixed with amateur or historical observatories mostly resided within human activities.
Therefore, statistical trends in Fig. \ref{F:hist} are mostly dominated by these so called ``outlier observatories''.
This fact was also concluded in \citep{2020MNRAS.493.1204A}.

\subsection{CC}
Most of the observatories are accumulated on the high CC value part of Fig. \ref{F:hist} corresponding to a ``cloudy'' weather.
The mean of CC is $\sim$70\% which corresponds to 110 clear nights per year and only 1.5\% of observatories (N=32) are above 3$\sigma$ cloud coverage i.e having larger number of clear nights.
Cerro Paranal, Chile is found to have the lowest CC value of 9.02\% which corresponds 333 clear nights per year.

\subsection{PWV}
PWV distribution of observatories displays a good Gaussian profile around the mean of $\sim$14 mm.
Only one observatory found within 3$\sigma$.
It is the driest site among the others: Corona Borealis Obs. (at 4965 m; site named as Ali in Tibet, China).
If it is extended to 2$\sigma$ then number of observatories increases to 26 corresponding to 1.2\%.

\subsection{AOD}
As can be observed in \ref{F:hist}, AOD is equally distributed around the mean value of 1.81.
Therefore, it is easy to find a statistical trend for this layer.
However, only one observatory found within 2$\sigma$: South African Astronomical Obs., Northern Cape/Africa.

\subsection{DEM}
Observatories are mostly populated to low DEM values.
Almost 70\% of observatories (N=1502) are below the mean of DEM (550 m) and only $\sim$2\% of observatories (N=37) are above 3$\sigma$ elevation.
Therefore, observatories above 3$\sigma$ (N=37) can be counted as professional and located in high altitudes.
Among 1$\sigma$ the highest DEM point refers to Corona Borealis Obs.(at 4965 m; site named as Ali in Tibet, China).

\subsection{HWV}
The statistical distribution of this parameter is uneven and the $\sigma$ value is too high.
However, there is a narrow peak at around 15 m/s.
Since the data is not distributed around this peak, beyond the peak (19-45 m/s) it is hard to find a statistical significance.
Note that ``low astronomical seeing'' should have a low value of HWV which is seen below 12 m/s.
There are 17 observatories falling into this portion of the distribution and only one observatory is below 3 m/s: Observatorio Astronomico UTP Pereira, Colombia.

\subsection{VWV}
The best observatories should have VWV value close to zero.
VWV distribution of observatories displays a good Gaussian profile around a sharp mean of $\sim$0.04 Pa/s.
Since VWV values are distributed around zero, there are in total 430 observatories within $\pm1\sigma$, corresponding to $\sim$20\%.
Since most of observatories have low vertical velocities (VWV) one cannot distinguish them around VWV=0.0.


\section{Results of Monthly and Annual Averages}

In order to have a global outcome of astronomical weather changes though out the time span of our dataset a subset of observatories listed in Table \ref{T:factor_ave} was selected.
Therefore, for these observatories monthly and annual (Fig. \ref{F:north} and Fig. \ref{F:south}) averages are tabulated and plotted.
In order to have a quick look to the status of the observatories for all factors, average of annual changes are tabulated in Table \ref{T:factor_ave}.
The observatories are divided into two major groups according to their geographic locations, namely hemisphere of the observatory location.

There is a clear difference between both hemispheres.
Even though number of southern observatories are less compared to northern ones, the changes (both monthly and yearly) are more abrupt and chaotic for northern observatories whereas southern observatories show similar trends and changes.

When both monthly and annular changes are considered an expected outcome would be summarized as the following:
If observatory's monthly changes is high/low then it's annual change stays high/low too.
In addition to this, trend of annual changes stay almost constant through out the time span of the dataset except for some of the outlier observatories.

\subsection{CC}
All northern observatories show disordered trend in monthly CC changes. However, there is a clear and obvious trend for southern observatories.
All southern observatories annually stay below around 38\% CC value.

Minimum CC value is 19\% (Jun) for PalO in northern hemisphere, and 3\% (Mar and Nov) for CA and PO in southern hemisphere.
The worst CC values (50-95\%) are reached for most of the northern observatories.
Northern observatories could also be grouped into 3 `annual' percentiles: 55\%, 65\% and 75\%.

Excluding outlier observatories northern and southern observatories have reversed peaks: November to March for northern hemisphere, and May to June for southern hemisphere.
Reversal of these peaks could easily be correlated to seasonal (and/or climatic) changes of the corresponding hemisphere.

\subsection{PWV}
Hemispheres are reversed in trends of PWV changes:
Monthly PWV averages peak around July-August for northern hemisphere whereas they peak around January-February for southern hemisphere.

Lower background PWV level stays around 6 mm for both hemispheres from November to April (6 months) for northern hemisphere and June to October (5 months) for southern hemisphere.

All southern observatories stay below 15 mm PWV value whereas most of northern observatories reach to around 30 mm PWV value.
The highest PWV values of MKO and ORM (around 17 mm) make them outliers for both monthly and annual changes.
The lowest level of PWV change (5.5 mm averaged over whole dataset) is observed for DAG.

\subsection{AOD}
There are many outlier observatories in monthly averages of AOD: BAO, MRO, ORM and SAO in northern hemisphere, and PO in southern hemisphere. Since DAG's AOD values for Dec to Feb are missing, and have high values for Mar to Apr, DAG is also marked as outlier.
When these outliers are excluded rest of the observatories stay below 1.0 AOD value.

The sharp peaks in annular changes for southern hemisphere probably corresponding to biomass burning activity \citep{2010ACP....10.3505T}.
There is a slight increase in July and August for both hemispheres.

\subsection{HWV}
Hemispheres are reversed in trends of HWV changes:
Monthly HWV averages peak around 27 m/s (corresponding to 22 m/s annually) from May to October whereas they peak around 37 m/s (corresponding to 32 m/s annually) from May to November.

BAO can be marked as outlier due to its high overall monthly change (min-max: 24-38 m/s).
The lowest HWV value is around 6 m/s in July and August for southern observatories of MDO, FLWO, KPNO and MGIO:
All of the observatories stay below 40 m/s through out the year.

HWV values increases unexpectedly from June to September for DAG, SAO and MRO.

\subsection{VWV}
Hemispheres show different trends of VWV changes.
Monthly change of VWV for all southern observatories are found to be upward and steady, excluding the outlier CTIO.
This  can be also confirmed on the annual trend graph.

Two observatories (CA and LCO) levels up from Apr to Aug.

Looking first to the annual changes of northern observatories one could easily find out two distinguishing groups of observatories: 1) Having downward VWV level (MDO, FLWO, KPNO, MGIO, PalO, LO, MRO); 2) Having upward VWV level (MKO, ORM, DAG, BAO, SAO).
These groups show similar trends in the monthly changes.
Latter group of observatories reach to maximum downward VWV in Jun.

The minimum VWV variation which corresponds to the most stable atmosphere above the observatory, is found in ORM (northern) and PO (southern).

\section{Results of Correlations}

\begin{figure*}
    \centering
    \begin{tabular}{@{}c@{}c@{}}
    \includegraphics[height=5.25cm]{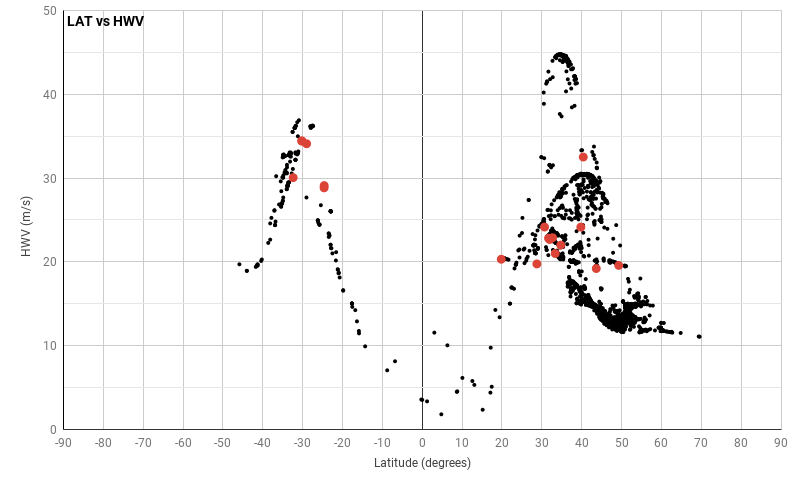}    &
    \includegraphics[height=5.25cm]{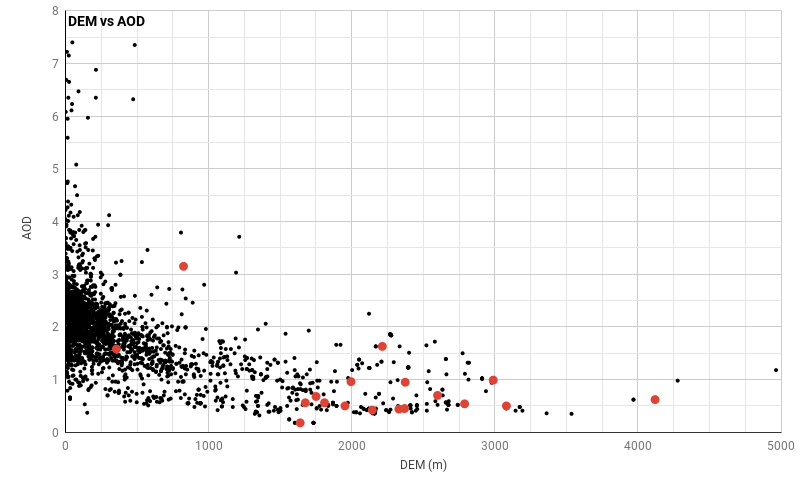}    \\
    \includegraphics[height=5.25cm]{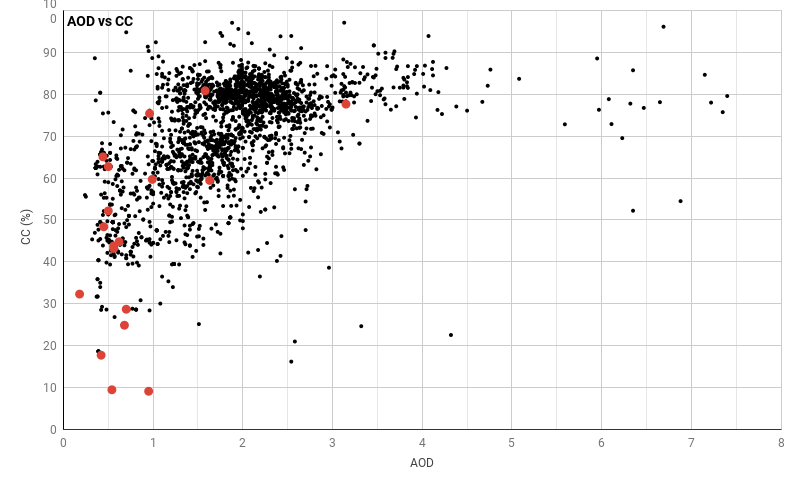}     &
    \includegraphics[height=5.25cm]{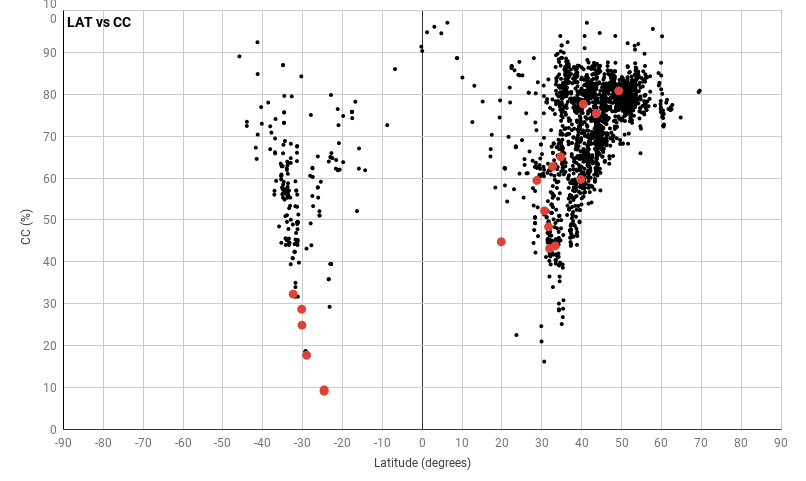}     \\
    \includegraphics[height=5.25cm]{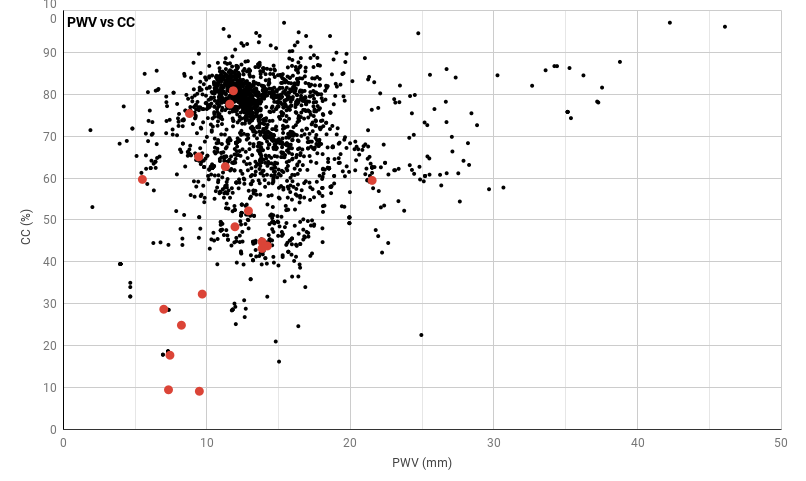}    &
    \includegraphics[height=5.25cm]{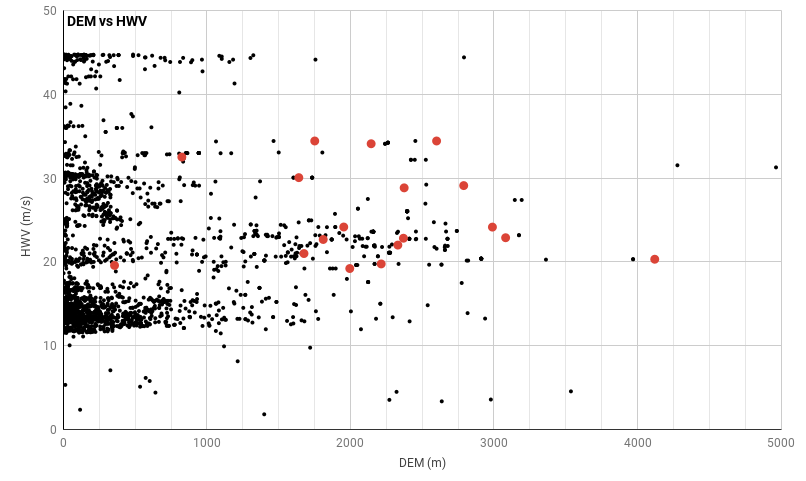}   \\
    \includegraphics[height=5.25cm]{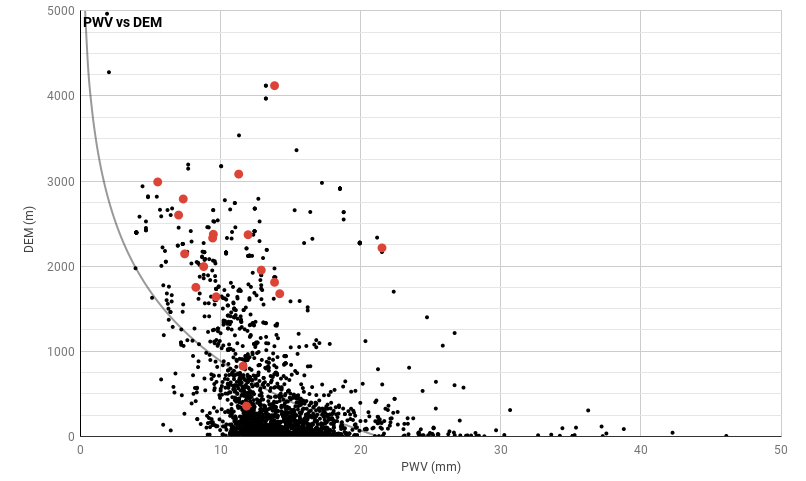}    &
    \includegraphics[height=5.25cm]{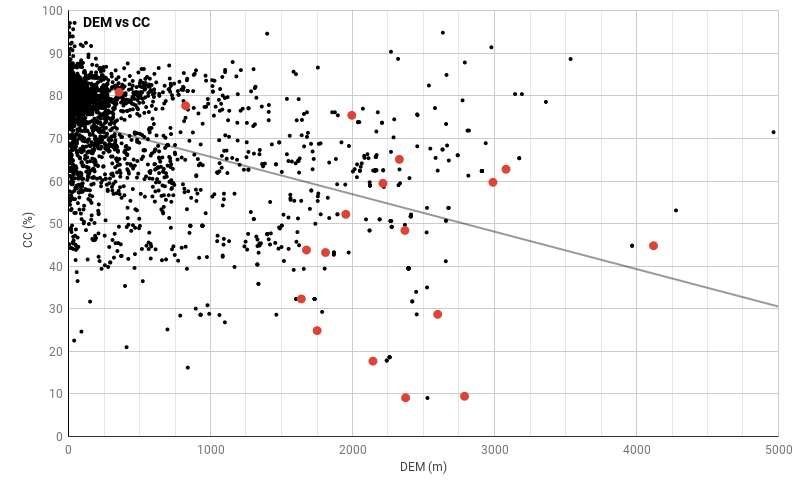}   \\
    \end{tabular}
\caption{%
For each observatory we plot the cross-correlations between certain layers given in Table \ref{T:coef} (4 graph in top two panels).
Black points represent the full list of observatories and red points are short list large-sized observatories.
The third panel contains PWV vs CC, and DEM vs HWV and they are given as an example to \textbf{uncorrelated} layers.
Special correlations are given in the bottom panel: between PWV and DEM (left), and between DEM and CC (right).
A logarithmic fit drawn for PWV--DEM and a linear fit drawn for DEM--CC is shown with solid black line.%
}
\label{F:corr_new}
\end{figure*}

The correlations between the factors are given in Table \ref{T:coef} and explained in \S\ref{sec:Correlations} and their related graphs are given in Fig. \ref{F:corr_new}.

There is a special condition for VWV.
It has two directions: upward (positive) and downward (negative).
The goodness of correlations for combined VWV gives the following correlation results: DEM-VWV (-0.89) and VWV-AOD (0.64).
In order to assess goodness of correlations correctly, VWV is divided into two components: VWVu (upward) and VWVd (downward), and it is given in the lower panel of Table \ref{T:coef}.
Therefore, abovementioned correlations of DEM-VWV and VWV-AOD coudl easily be ignored.

The results of other correlations are discussed below:

\textbf{LAT-HWV:} As LAT increases HWV increases.
For northern observatories, HWV peaks at $\phi\sim35\degr$ and $\phi\sim40\degr$ latitudes, however for southern observations, HWV peaks only at $\phi\sim-30\degr$ latitude.
This correlation shows almost a reversed pattern of LAT-CC correlation.

\textbf{DEM-AOD:} As DEM increases AOD decreases.
This is an expected relation between two parameters from the astronomical point of view.
The trend might be thought as exponential.
The relation can be divided at AOD value 1.0:
for most of observatories AOD is above this value and 
for most of the higher observatories (mostly red dots) AOD stays below this value.

\textbf{AOD-CC:} As CC increases AOD increases.
This is an expected relation between two parameters from the astronomical point of view.
The trend might be thought as quadratic.
AOD value of 2.0 is concentrated around CC value of 80\%.
The selected higher altitude observatories have low CC value therefore their AOD value stays below 1.0.

\textbf{LAT-CC:} As LAT increases CC increases.
This is an expected relation between two parameters from the astronomical point of view.
Regardless of the hemispheres, CC reaches to minimum value for $|\phi|\sim30\degr$.

\textbf{PWV-CC:} shows no correlation.
The factors are distributed around a central value of PWV=12 mm and CC=80\%.
However, there is no clear trend due to this concentrated distribution.

\textbf{DEM-HWV:} shows no correlation.
The factors are distributed along several different quantised HWV values: 15 m/s, 20 m/s, 30 m/s and 45 m/s.
Therefore, one cannot deduce a common trend from these quantised HWV values.

\textbf{PWV-DEM:} As DEM increases PWV decreases.
This is an expected relation between two parameters from the astronomical point of view.
The trend might be thought as exponential: $DEM = 3585-1173\times\ln(PWV)$.
Using a different technique, \cite{2019PASP..131d5001O} found a similar exponential trend between these two factors. The relation was also shown by \cite{2019MNRAS.482.4941H}.
Most of the observatories are concentrated at PWV=15 mm and DEM below 1000 m.
Due to this concentration, the higher altitude observatories are not on the upper edge of exponential function, however, they clearly show the trend (red dots).

\textbf{DEM-CC:} A common expectation in astronomy is that \textit{as DEM increases CC decreases}  \citep{2019MNRAS.482.4941H, 2020MNRAS.493.1204A, 2015ExA....39..547A}. 
However, our Spearman statistics gives a \textbf{weak correlation} with -0.41 (see Fig. \ref{F:corr_new}, bottom right panel).
Note also that CC value of all observatories below DEM $<1000$ m clustered around 80\% and CC value of the short list of observatories stays below 70\%.
As a conclusion \textit{as DEM increases CC decreases} is real (see the fit in the graph) however, it cannot be taken as a full-proof concept in astronomy due to the worst fit with R=0.21 ($CC = -0.0088 \times DEM + 74.5$), therefore it has to be carefully considered by including other factors to the individual site values including also the local trends of the atmosphere above the site.

Above results have also been checked against an earlier study made by \citet{2019MNRAS.482.4941H}.
However, due to (a) having very different time spans of datasets, and (b) different methods and models used in the analysis of data, no clear and reliable comparison were produced.
\section{Conclusion}
\label{sec:Conclusion}

All cataloged worldwide observatories (2141 in total) were used to assess their atmospheric conditions from the point of view of astronomy.
The following factors were chosen to create GIS layers at their geographic locations from the \href{https://astrogis.org}{astroGIS database}: CC (cloud coverage), PWV (precipitable water vapor), AOD (atmospheric optical depth), VWV (vertical wind velocity) and HWV (horizontal wind velocity).
In order to estimate astronomical importance of their geographic locations and quality of airmass above the observatories, DEM (digital elevation model) and LAT (latitude of observatory location) layers were also included.
Two periodic variations have been produced from these factors: monthly and annual averages.
In addition to the variations or trends a complete statistical analysis was carried out for all factors to investigate the potential correlations between the factors.
A short list of 14 observatories is marked through out the study indicating ``observatory with a telescope of 4m+ in size''.
Note that a recent and similar study was carried out for a selected of 9 observatories by \cite{zuhal2021}.
All the results will be made available online through astroGIS database.

The global weather patterns effects all astronomical factors of this study.
Their monthly and/or annual changes, however, show various different effects on individual factors.
Therefore, it is easy to conclude that ``astronomical importance'' of a factor cannot be relied on itself only and it has to be shaped with all factors too.
The outcomes of this complete analysis are listed below:
\begin{itemize}
\item
There is a clear difference between northern (NH) and southern (SH) hemispheres for all factors.
SH seems to be calmer than NH in two ways: (a) overall values are lower, (b) annual variations are smaller.

\item
In order to take into account the seasonal variations from our datasets we defined two distinct seasons excluding the transitional spring periods: summer and winter.
The summer period for NH (SH) is from May to September (November to March), and the winter period is exchanged between NH and SH.
These periods are also compatible from the point of meteorology \citet{1983BAMS...64.1276T}.
The exchange of seasons between NH and SH are apparent in all factors, distinctively in PWV, CC, AOD and HWV.

\item
Our list contains 2141 individual observatory (1905 northern and 235 southern).
Since there is no official list maintained by international astronomical organizations the list is far from professional quality because observatories producing scientific data are mixed with amateur or historical observatories mostly resided within human activities.
We also confirm this fact in the outcomes of correlations for all factors, especially when the factors of DEM, CC and PWV are included.

Therefore, the geographical locations of most of the observatories found to be ``not suitable for astronomical observations'', confirming once again \cite{2020MNRAS.493.1204A}: \textit{only approximately 10 per cent of all current observatories are located in good locations in all SIAS series}.

A new observatory list will be recompiled and revised from the old one, and it will be published elsewhere.
Abovementioned outcomes will be reassessed with the new observatory list.


\item
\textbf{Confirmed:} No long-term trends and/or pattern in all factors.
Even though we present only annual averages for the short list of observatories the trend is similar for the whole list.
Note also that annual variation patterns continue from the monthly variations.

\item
\textbf{Confirmed:} No \textit{clear} correlations were found for PWV-CC and DEM-HWV.
This is somehow expected due to natures of PWV and HWV.
PWV-CC data is crowded with outlier observatories; when the data re-compiled for observatories having high DEM values (e.g. short list of our observatory database) they appear to give a higher correlations.
Similarly, since HWV has a natural quantized values at certain geographic locations it is not easy to have a ``clear trend''.

\item
\textbf{Confirmed:} As DEM increases astronomical conditions gets better with lower CC, PWV and AOD values (see Fig. \ref{F:corr_new} for the related correlations).
In order to understand the observatory conditions one could use the relations to estimate individual status of location.

Note that even though DEM-AOD correlation is high and clear, the correlations for PWV-DEM and DEM-CC is low and weak.
However, when PWV and CC are assisted with other site properties they usually confirm the validity of the correlation.

\end{itemize}

\section*{Acknowledgements}
This research was supported by the Scientific and Technological Research Council of Turkey (TÜBİTAK) through project number 117F309. This research was also supported by the Çukurova University Research Fund through project number FYL-2019-11834.

\section*{Data Availability}
This work has made use of astroGIS database released under the web site astrogis.org.
The data will be made available on request from the corresponding author.

\section*{Supplementary Material}
\label{sec:Supl}

Additional supporting information includes four files in the online version of this article.

\textbf{\texttt{Table-1\_Full.txt}}: Complete version of Table \ref{T:obs}; containing country, altitude and location astronomical sites.

\textbf{\texttt{Table-4\_Full.txt}}: Complete version of Table \ref{T:factor_ave}; average of annual averages of all astronomical sites for CC, PWV, AOD, VWV and HWV factors.

\textbf{\texttt{Figure-3,4\_Monthly-Full.txt}}: Monthly averages for all astronomical sites within time span of dataset for CC, PWV, AOD, VWV and HWV factors.

\textbf{\texttt{Figure-3,4\_Yearly-Full.txt}}: Yearly averages for all astronomical sites within time span of dataset for CC, PWV, AOD, VWV and HWV factors.

\bibliographystyle{mnras}
\bibliography{w-0} 

\begin{table*}
    \caption{%
    The locations of astronomical sites hosting a +4 m class telescope.
    The table is in decreasing latitude order from South to North.
    \underline{Observatory abbreviations} (given geographic coordinates usually represents a single telescope among others):
    SAAO: South African Astronomical Obs.;
    CPO: Cerro Pachón Obs.;
    CTIO: Cerro Tololo Inter-American Obs.;
    LCO: Las Campanas Obs.;
    CA: Cerro Armazones.;
    PO: Paranal Obs.;
    MKO: Mauna Kea Obs.;
    ORM: Roque de los Muchachos Obs.;
    MDO: McDonald Obs.;
    FLWO: Fred Lawrence Whipple Obs.;    
    KPNO: Kitt Peak National Obs.;
    MGIO: Mount Graham International Obs.;
    PalO: Palomar Obs.;
    LO: Lowell Obs.;
    DAG: Eastern Anatolia Obs.;
    BAO: Beijing Astronomical Obs.;
    SAO: Special Astrophysical Obs.;
    MRO: Maple Ridge Obs.
    Full observatory list will be available online.%
}
    \centering
    \begin{tabular}{lcccc}
    \hline
    Site & Country & Altitude & Longitude & Latitude \\
& & (m) & (deg)& (deg) \\
    \hline
    SAAO &Africa& 1733 & 20.81  & -32.37 \\
    CPO  &Chile& 2452 & -70.73  & -30.24 \\
    CTIO &Chile& 1751 & -70.80  & -30.16 \\
    LCO  &Chile& 2144 & -70.70  & -29.00 \\
    CA   &Chile& 2789 & -70.20  & -24.60 \\
    PO   &Chile& 2374 & -70.40  & -24.62 \\
    MKO  &USA& 4120 & -155.47 & 19.82  \\
    ORM  &Spain& 2214 & -17.89  & 28.75  \\
    MDO  &USA& 1894 & -104.02  & 30.67  \\
    USA$^{*}$ &USA& 1810 & -111.59 & 31.95 \\
    DAG  &Turkey& 2989 & 41.23   & 39.78  \\
    BAO  &China& 825  & 117.57  & 40.39  \\
    SAO  &Russia& 1897 & 41.44   & 43.64  \\
    MRO  &Canada& 365  & -122.57 & 49.28  \\
    \hline
    \end{tabular}
    \label{T:obs}
    \\ \footnotesize{$^{*}$ FLWO, KPNO, MGIO, PalO and LO observatories in USA are labeled as USA (see Fig. \ref{F:obs}).}
\end{table*}
\begin{table*}
    \centering
    \caption{%
    The properties of layers used in astroGIS database.
    }
    \begin{tabular}{l c c cc c}
    \hline
Layer
    & Coverage
    & \multicolumn{1}{c}{Resolutions}
    & \multicolumn{1}{c}{Unit}
    & \multicolumn{2}{C{3cm}}{Scale Definitions} \\
    & 
    & \multicolumn{1}{c}{km}
    &
    & \multicolumn{1}{c}{Worst}
    & \multicolumn{1}{c}{Best} \\
    \hline
CC
    &2000-2019
    & 3
    & \%
    & High & Low \\
PWV
    &2000-2019
    & 5
    & mm
    & High & Low\\
AOD
    &2000-2019
    & 10
    & -
    & High & Low\\
DEM
    & 1996
    & 1  
    & m
    & Low & High \\
VWV*
    & 1979-2019
    & 31 
    & Pa s$^{-1}$
    & Fast & Slow\\ 
HWV*
    & 1979-2019
    & 31  
    & m s$^{-1}$ 
    & Fast & Slow\\
    \hline
    \end{tabular}
    \\ \footnotesize{* Layer data has been adapted from the ERA5 database.}
\label{T:astrogis}
 \end{table*}
\begin{table*}
    \caption{%
    The layer statistics and layer percentiles of the observatories.
    $\mu$(1), $\mu$(2) and $\mu$(3) are given in counts (N) and percentages (\%) of observatories falling into 1, 2 and 3 sigma, respectively, centered on the mean of the layer.
    Similarly, $\ge\mu$ and $<\mu$ (in counts and percentages) show the full distribution statistics.
    The sign of the sigma values are; for DEM `$+$', for VWV `$\pm$', for the rest the factors sigma is `$-$'.%
}
    \begin{tabular}{@{}c cccc @{~}rrrr
            rrrrrr@{}}
    \cline{2-15}
    & \multicolumn{4}{c@{}}{Statistics of the layers}
    & \multicolumn{10}{c@{}}{Observatories (N) \& Ratio (\%)} \\
    \hline
\multirow{2}{*}{Layer}
    & \multirow{2}{*}{Min}
    & \multirow{2}{*}{Mean}
    & \multirow{2}{*}{Max}
    & \multirow{2}{*}{Sigma}
    & \multicolumn{2}{C{15mm}}{$\ge\mu$}
    & \multicolumn{2}{C{15mm}}{$<\mu$}
    & \multicolumn{2}{C{15mm}}{$\mu$(1)}
    & \multicolumn{2}{C{15mm}}{$\mu$(2)}
    & \multicolumn{2}{C{15mm}@{}}{$\mu$(3)}\\
    \cline{6-15}

    & 
    & 
    & 
    & 
    & (N) & (\%)
    & (N) & (\%)
    & (N) & (\%)
    & (N) & (\%)
    & (N) & (\%) \\
    \hline
CC
    & 9.02
    & 69.65
    & 97.15
    & 14.31
    & 1256
    & 58.6
    & 885
    & 41.3
    & 327
    & 15.3
    & 91
    & 4.2
    & 32
    & 1.5\\
    PWV
    & 1.88
    & 13.85
    & 46.09
    & 4.16
    & 882
    & 41.2
    & 1259
    & 58.8
    & 206
    & 9.6
    & 26
    & 1.2
    & -
    & 0.0\\
    AOD
    & 0.21
    & 1.81
    & 7.17
    & 0.88
    & 1081
    & 50.05
    & 1060
    & 49.5
    & 324
    & 15.1
    & -
    & 0.0
    & -
    & 0.0\\
     DEM
    & -63*
    & 550
    & 4965
    & 739.97
    & 639
    & 29.8
    & 1502
    & 70.1
    & 325
    & 15.2
    & 165
    & 7.7
    & 37
    & 1.7\\
    \multirow{2}{*}{VWV}
    & \multirow{2}{*}{-0.42}
    & \multirow{2}{*}{0.04}
    & \multirow{2}{*}{0.89}
    & \multirow{2}{*}{0.12}
    & \multirow{2}{*}{802}
    & \multirow{2}{*}{37.5}
    & \multirow{2}{*}{1339}
    & \multirow{2}{*}{62.5}
    & 158
    & 7.4
    & 58
    & 2.7
    & 15
    & 0.7\\
    & 
    & 
    & 
    & 
    & 
    & 
    & 
    & 
    & 272
    & 12.7
    & 89
    & 4.2
    & 23
    & 1.1\\
     HWV
    & 1.78
    & 21.06
    & 44.79
    & 9.02
    & 864
    & 40.4
    & 1277
    & 59.6
    & 79
    & 3.7
    & 2
    & 0.1
    & -
    & 0.0\\
    \hline
    \end{tabular}
    \label{T:layer_stat}
    \\ \footnotesize{*This data come from Digital Elevation Model that is not real elevation of the observatory.}
\end{table*}
\begin{table*}
    \caption{%
    The average of annual averages for CC, PWV, AOD, VWV and HWV factors.
    The average values are given along with their standard deviations.
    Full observatory list will be available online.%
    }
    \centering
    \begin{tabular}{lrrrrr}
    \hline
Site
& \multicolumn{1}{c}{CC}
& \multicolumn{1}{c}{PWV}
& \multicolumn{1}{c}{AOD}
& \multicolumn{1}{c}{VWV}
& \multicolumn{1}{c}{HWV}
\\ 

& \multicolumn{1}{c}{\%}
& \multicolumn{1}{c}{mm}
& \multicolumn{1}{c}{value}
& \multicolumn{1}{c}{Pa s$^{-1}$}
& \multicolumn{1}{c}{m s$^{-1}$}
\\ 

\hline

SAAO &	32.29	$\pm$ 	2.91    &	9.67	$\pm$ 	0.47	&	0.21	$\pm$	0.01	&	0.33	$\pm$ 	0.02	&	30.06	$\pm$ 	1.77	\\
CPO &	28.67	$\pm$ 	3.70    &	6.99	$\pm$ 	0.71	&	0.48	$\pm$	0.14	&	0.30	$\pm$ 	0.02	&	34.44	$\pm$ 	1.93	\\
CTIO &	24.85	$\pm$ 	3.86	&	8.22	$\pm$ 	0.83	&	0.36	$\pm$	0.07	&	0.39	$\pm$ 	0.02	&	34.44	$\pm$ 	1.93	\\
LCO	&	17.68	$\pm$ 	3.59	&	7.42	$\pm$ 	0.99	&	0.36	$\pm$	0.05	&	0.14	$\pm$ 	0.01	&	34.11	$\pm$ 	2.04	\\
CA	&	9.42	$\pm$ 	2.12	&	7.32	$\pm$ 	0.52	&	0.54	$\pm$	0.06	&	0.12	$\pm$ 	0.03	&	29.11	$\pm$ 	2.29	\\
PO	&	9.07	$\pm$ 	1.90    &	9.47	$\pm$ 	0.62	&	0.95	$\pm$	0.06	&	0.30	$\pm$ 	0.02	&	28.84	$\pm$ 	2.26	\\
MKO	&	44.79	$\pm$ 	5.56	&	13.84	$\pm$ 	1.72	&	0.67	$\pm$	0.15	&	0.24	$\pm$ 	0.04	&	20.31	$\pm$ 	2.79	\\
ORM	&	59.46	$\pm$ 	3.85	&	21.51	$\pm$ 	1.24	&	1.58	$\pm$	0.17	&	0.15	$\pm$ 	0.02	&	19.75	$\pm$ 	1.81	\\
MDO	&	52.18	$\pm$ 	5.23	&	12.89	$\pm$ 	0.83	&	0.42	$\pm$	0.04	&	-0.03	$\pm$ 	0.03	&	24.16	$\pm$ 	1.77	\\
USA	&	43.20	$\pm$ 	2.60	&	13.84	$\pm$ 	0.68	&	0.40	$\pm$	0.05	&	-0.26	$\pm$ 	0.03	&	22.67	$\pm$ 	2.19	\\
DAG	&	59.70	$\pm$ 	3.05	&	5.50    $\pm$ 	0.48	&	0.90	$\pm$	0.15	&	0.10	$\pm$ 	0.01	&	24.15	$\pm$ 	1.43	\\
BAO	&	77.69	$\pm$ 	1.91	&	11.59	$\pm$ 	0.73	&	3.20	$\pm$	0.66	&	0.14	$\pm$ 	0.01	&	32.52	$\pm$ 	1.92	\\
SAO	&	75.44	$\pm$ 	3.62	&	8.78	$\pm$ 	0.83	&	0.98	$\pm$	0.17	&	0.26	$\pm$ 	0.04	&	19.20	$\pm$ 	1.41	\\
MRO	&	80.87	$\pm$ 	2.86	&	11.84	$\pm$ 	0.62	&	1.63	$\pm$	0.68	&	-0.11	$\pm$ 	0.03	&	19.57	$\pm$ 	1.73	\\

    \hline
    \end{tabular}
    \label{T:factor_ave}
\end{table*}
\bgroup
\renewcommand{\arraystretch}{0.75}
\setlength{\fboxsep}{2.5pt}
\setlength{\fboxrule}{1pt} 
\setlength{\tabcolsep}{0pt}
\setlength\extrarowheight{3mm}
\begin{table*}
    \caption{%
Spearman({$\rho$}), Pearson (r) and Kendall({$\tau$}) correlation coefficients for average annual changes of all the layers.
\textit{Latitude} of the observatory site (aka. LAT) was also included as another correlation variable.
Color codes: \colorbox{amber}{Amber} color grading represents direct correlation with numbers ranging from 0.00 to 1.00 (max); \colorbox{aqua}{Aqua} color grading represents inverse correlation with numbers ranging from 0.00 to -1.00 (max).
Note that since VWV has two opposite directions (upward and downward) their statistics have to be analyzed accordingly.
This is given in the lower panel, however only for Spearman since Pearson and Kendall show similar trends on the upper panel.
}
\centering
\begin{small}
\begin{tabular}{@{}c ZZZZZZZZZZZZZZZZZZZZZZZZ  @{}}
\hline
    & \multicolumn{7}{c}{Spearman ($\rho$)}
    & \multicolumn{7}{c}{Pearson (r)}
    & \multicolumn{7}{c}{Kendall ($\tau$)}\\
    & \multicolumn{1}{c}{CC} & \multicolumn{1}{c}{PWV}  & \multicolumn{1}{c}{AOD}
    & \multicolumn{1}{c}{VWV} & \multicolumn{1}{c}{HWV} & \multicolumn{1}{c}{DEM}
    & \multicolumn{1}{c}{LAT}
    & \multicolumn{1}{c}{CC} & \multicolumn{1}{c}{PWV}  & \multicolumn{1}{c}{AOD}
    & \multicolumn{1}{c}{VWV} & \multicolumn{1}{c}{HWV} & \multicolumn{1}{c}{DEM}
    & \multicolumn{1}{c}{LAT}
    & \multicolumn{1}{c}{CC} & \multicolumn{1}{c}{PWV}  & \multicolumn{1}{c}{AOD}
    & \multicolumn{1}{c}{VWV} & \multicolumn{1}{c}{HWV} & \multicolumn{1}{c}{DEM}
    & \multicolumn{1}{c}{LAT} \\
\noalign{\smallskip}
\hline
CC  &  1.00	& 	& 	& 	& 	& 	&
    &  1.00	& 	& 	& 	& 	& 	&
    &  1.00	& 	& 	& 	& 	& 	& 	\\
PWV & -0.06	&  1.00	& 	& 	& 	&   &
    &  0.07	&  1.00	& 	& 	& 	&   &
    & -0.05	&  1.00	& 	& 	& 	& 	& 	\\
AOD &  0.57	& 0.25	&  1.00	& 	& 	&  	&
    &  0.53	&  0.34	&  1.00	& 	& 	&   & 
    &  0.40	&  0.17	&  1.00	& 	& 	& 	& 	\\
VWV &  0.40	& 0.39	& 0.64	&  1.00	& 	&   & 	
    &  0.45	& 0.36	& 0.57	&  1.00	& 	&   &
    &  0.27	& 0.27	& 0.46	&  1.00	& 	& 	& 	\\
HWV & -0.20	&  0.32	& -0.19	& -0.11	&  1.00	&   &  	
    & -0.12	&  0.16	&  0.11	& -0.06	&  1.00	&   &
    & -0.16	&  0.22	&  0.15	& -0.07	&  1.00	& 	& 	\\
DEM & -0.41	& -0.43	& -0.66	& -0.89 &  0.08	&  1.00 &	
    & -0.44	& -0.36	& -0.56	& -0.99	&  0.05	&  1.00	&
    & -0.28	& -0.30	& -0.48	& -0.73	&  0.05	&  1.00	& 	\\
LAT &  0.54	& -0.41	&  0.45	& 0.34	& -0.70	& -0.33	&  1.00
    &  0.50	& -0.47	&  0.25	& 0.43	& -0.47	& -0.43	&  1.00
    &  0.38	& -0.30	&  0.33	& 0.24	& -0.51	& -0.23	&  1.00	\\
%
%
%
%
%
%
    \noalign{\smallskip}
    \hline
    & \multicolumn{1}{c}{CC}  & \multicolumn{1}{c}{PWV} & \multicolumn{1}{c}{AOD}
    & 
    & \multicolumn{1}{c}{HWV} & \multicolumn{1}{c}{DEM} & \multicolumn{1}{c}{LAT} 
    &  
    & & & & & & & & & & & & & \\
VWVu& -0.21	& -0.24	& -0.36	& & 0.01  & 0.42  & 0.39  & 
    & & & & & & & & & & & & & \\
VWVd& 0.38	& 0.08	& 0.47	& & -0.26 & -0.48 & -0.23  & 
    & & & & & & & & & & & & & \\
    \noalign{\smallskip}
    \hline
    \end{tabular}
    \end{small}
    \label{T:coef}
\end{table*}
\egroup

\bsp	
\label{lastpage}
\end{document}